\documentclass[
  aps,prb,twocolumn,showpacs,amsmath,amssymb,superscriptaddress
]{revtex4-1}

\makeatletter
\newcommand{\whencolumns}[2]{\preprintsty@sw{#1}{#2}}
\makeatother

\usepackage[english]{babel}
\usepackage{amsmath,amssymb,amsfonts} 	
\usepackage{graphicx}
\usepackage{array,multirow}
\usepackage[utf8]{inputenc}
\usepackage{bm}
\usepackage{xcolor}
\usepackage[normalem]{ulem}
\usepackage{siunitx}
\usepackage{hyperref}
\usepackage{xspace}
\hypersetup{colorlinks=true,linkcolor=blue,citecolor=blue,urlcolor=blue}
\usepackage{url}
\usepackage{textcomp}
\usepackage{chemformula}
\usepackage{comment}
\usepackage{lineno,hyperref}
\usepackage{float}
\usepackage{textcomp} 
\usepackage{rotating}
\usepackage[figurename=FIG.]{caption}
\usepackage{subcaption}
\usepackage{color} 
\definecolor{red}{rgb}{1.,0.,0.}
\usepackage{soul}
\usepackage{amsmath}
\usepackage{enumitem}

 \usepackage[
 justification=raggedright]
 {caption}
\usepackage{subcaption}
\usepackage{qcircuit}
\usepackage{braket}

\makeatletter
\newcommand*{\rom}[1]{\expandafter\@slowromancap\romannumeral #1@}
\makeatother

\newcommand\Harvard{ John A. Paulson School of Engineering and Applied Sciences, Harvard University, Cambridge, MA 02138, USA}
\newcommand\Sapienza{Department of Physics, Sapienza University of Rome, Piazzale Aldo Moro 5, 00185 Rome, Italy }
\newcommand\Bosch{Robert Bosch LLC Research and Technology Center, Watertown, MA 02472, USA}

\DeclareMathAlphabet\mathbfcal{OMS}{cmsy}{b}{n}

\newcommand{\mR}{\mathbfcal{R}}
\newcommand{\mP}{\mathbfcal{P}}
\newcommand{\bA}{\mathbf{A}}
\newcommand{\bB}{\mathbf{B}}
\newcommand{\bG}{\bm{\Gamma}}
\newcommand{\bR}{\mathbf{R}}

\newcommand{\dV}{\langle\mathbf{\partial^2V}\rangle}

\newcommand{\TOo}{{\ch{TO_1}}\xspace}
\newcommand{\TOd}{{\ch{TO_2}}\xspace}
\newcommand{\TOt}{{\ch{TO_3}}\xspace}
\newcommand{\TOf}{{\ch{TO_4}}\xspace}

\begin{document}

\title{Ultrafast quantum dynamics in $\mathbf{\mathrm{SrTiO_3}}$ under impulsive THz radiation}

\author{Francesco Libbi}
\thanks{Corresponding authors: \textcolor{blue}{libbi@g.harvard.edu}, \textcolor{blue}{lorenzo.monacelli@uniroma1.it}}
\author{Anders Johansson}
\affiliation{\Harvard}
\author{Boris Kozinsky}
\affiliation{\Harvard}
\affiliation{\Bosch}
\author{Lorenzo Monacelli}
\thanks{Corresponding authors: \textcolor{blue}{libbi@g.harvard.edu}, \textcolor{blue}{lorenzo.monacelli@uniroma1.it}}
\affiliation{\Sapienza}

\begin{abstract}
Ultrafast spectroscopy paved the way for probing transient states of matter produced through photoexcitation. Despite significant advances, the microscopic processes governing the formation of these states remain largely unknown. 
This study discloses the nuclear quantum dynamics of \ch{SrTiO3} when excited by THz laser pumping. 
We use a first-principles machine-learning approach accounting for all atomistic degrees of freedom to examine the time-resolved energy flow across phonon modes following the photoexcitation, revealing the mechanism underpinning the observed phonon upconversion and quantifying the lifetime of the out-of-equilibrium motion.
Crucially, our simulations predict that THz pump pulses can generate persistent out-of-equilibrium stress capable of inducing polar order. We observe a correlation between the measured lifetime of the transient inversion-symmetry-broken state and the duration of the out-of-equilibrium nuclear state.  
This work not only explains the experimental results on \ch{SrTiO3} but also establishes a framework for simulating the photoexcited quantum dynamics of nuclei from first principles without any empirical input. It lays the groundwork for systematic explorations of complex materials sensitive to photoexcitation.
\end{abstract}

\maketitle

Advances in ultrafast laser technology enable the generation of pulses that last for one electric field oscillation, providing direct probe of the out-of-equilibrium quantum dynamics of ions.
By employing ultra-short laser pulses to excite (pump) and subsequently probe the material's response, it is possible to capture transient dynamics, shedding light on new states of matter\cite{Disa2021} and allowing the manipulation of macroscopic properties such as polarization, magnetism, and conductivity. Dynamical switching between different states of matter like metal-insulator\cite{becker_femtosecond_1994,cavalleri_femtosecond_2001}, para-ferroelectric\cite{doi:10.1126/science.aaw4913,doi:10.1126/science.aaw4911}, para-ferromagnetic\cite{Disa2020}, topological\cite{sie_ultrafast_2019,mocatti_light-induced_2023} and even superconducting\cite{doi:10.1126/science.1197294} have been investigated, pointing to new possibilities for material design and dynamic control and promising transformational advances in technology and application.
On the other hand, intense pulses drive the system out of equilibrium \cite{10.1063/1.4983153}, illuminating the intrinsic interactions among collective atomic motions by probing the resulting nonlinear dynamics\cite{Kozina2019, Zhang2024, Zhang2024_1,Basini2024,PhysRevX.14.021016,PhysRevB.107.224307}.
Despite the rapid progress of the field, 
many of the mechanisms of the aforementioned phenomena are still actively debated. In fact, experimental probes only show a trace of the underlying complex quantum dynamics of ions and electrons, and a deeper theoretical understanding is needed, via computations of realistic materials' response to experimentally relevant stimuli.

\begin{figure*}
    \centering
    \includegraphics[width=\textwidth]{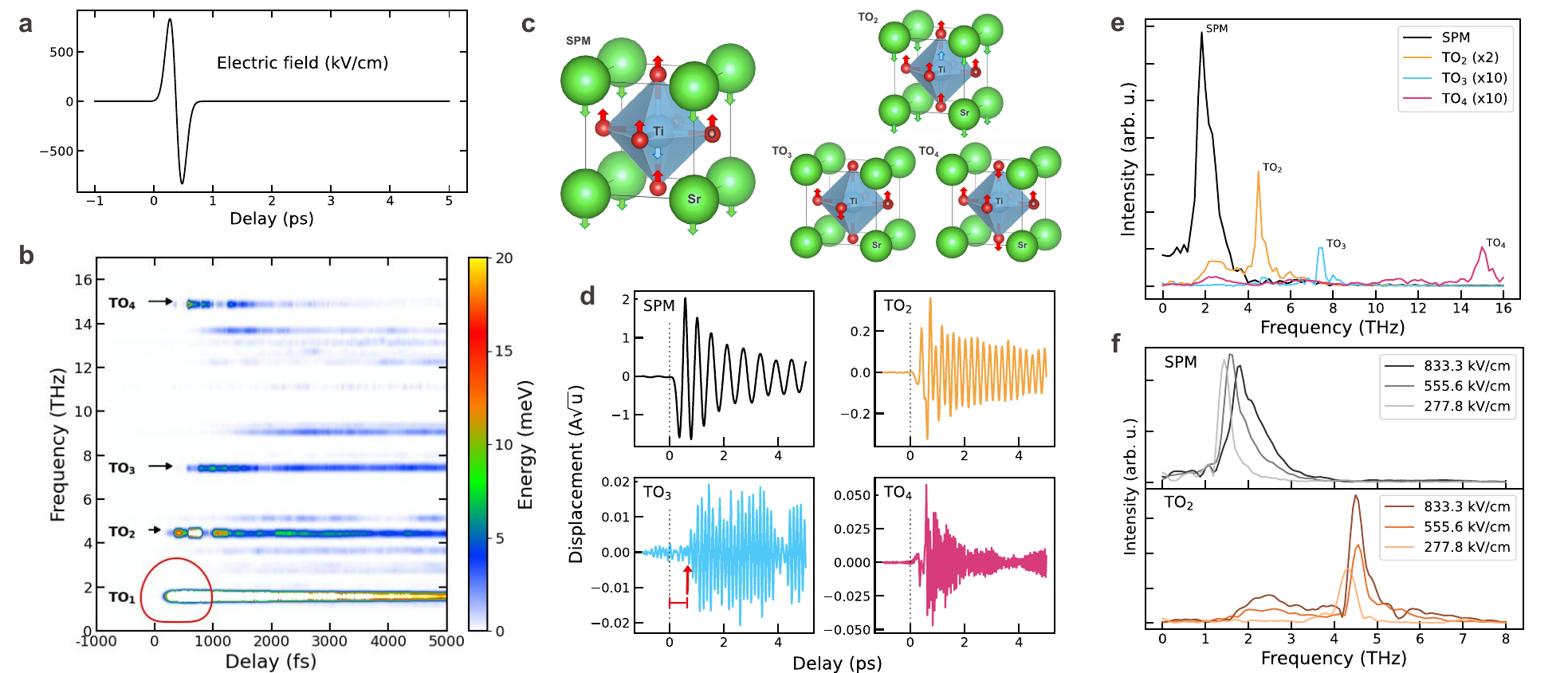}
    \caption{\textbf{a}, Pulse applied to STO. \textbf{b}, Phonon energy as a function of time and its natural frequency. \textbf{c}, Schematic representation of the 4 groups of acoustic modes at $\mathrm{\Gamma}$. \textbf{d}, $\mathrm{\Gamma}$-phonon displacement as a function of the pulse delay. \textbf{e}, Phonon spectral density, obtained as the Fourier transform of the displacements reported in Panel \textbf{d}. \textbf{f}, Dependence of the phonon spectral density as a function of the pulse amplitude; the signals have been zero-padded and windowed for processing.} 
    \label{fig1}
\end{figure*}

An outstanding example is the dynamic response to laser pulses of the \ch{SrTiO3} (STO) perovskite. In this material, nuclear quantum fluctuations are observed to suppress the ferroelectric order below \SI{37}{\kelvin} \cite{PhysRev.177.858}. This effect is due to the sufficiently small energy barrier of the double-well potential energy surface associated with the ferroelectric soft phonon mode (SPM), which allows the nuclear wavefunction to span the two minima, leading to tunneling and a consequent non-ferroelectric phase, stabilized by quantum fluctuations down to \SI{0}{\kelvin}\cite{PhysRevB.104.L060103, PhysRevMaterials.7.L030801}. 
Recent experiments reported the occurrence of second-harmonic generation (SHG) lasting for at least several picoseconds after an ultrafast excitation with THz light\cite{doi:10.1126/science.aaw4911,doi:10.1126/science.aaw4913}, which is a signature of the inversion symmetry breaking. This was interpreted as the formation of a ferroelectric metastable phase following the pump pulse\cite{PhysRevLett.129.167401}. However, this interpretation has been challenged by a similar observation on \ch{KTaO3}, another quantum paraelectric\cite{Ranalli2023}, where the appearance of the SHG after the pump was associated with the THz-driven dipole correlations of defect-induced local polar structures\cite{PhysRevLett.130.126902}. The nature of the light-induced SHG is therefore still highly debated.

The theoretical first-principles modeling of the light-induced response is necessary to examine the possible transient ferroelectric transition, but it is highly challenging. The current state-of-the-art approach is to build a low-dimensional model of the system retaining only a few degrees of freedom (the photoexcited vibrational mode and those modes deemed responsible for the occurrence of the light-induced transition), derive their anharmonic coupling from perturbation theory, and numerically determine the quantum dynamics of the model\cite{PhysRevLett.129.167401}. While this approach provides fundamental physical insight into the role of the coupling between the acoustic modes and the SPM to examine the light-induced ferroelectricity, it naturally misses important features of the out-of-equilibrium nuclear dynamics. The reduced size of these models, combined with the limited phonon-phonon scattering channels, can result in the emergence of fictitious metastable states and alter dissipation pathways, ultimately impacting the system's dynamics. In particular, real-time X-ray spectroscopy showed how photoexciting the SPM led to a complex energy transfer to high energy transverse optical (TO) modes on a femtosecond time scale. This phonon upconversion showcases how the real STO dynamics involve multiple degrees of freedom for which a full atomistic description is necessary\cite{Kozina2019}. 

One approach to study the atomistic quantum nuclear dynamics from first principles consists in the real-time path integral formulations \cite{10.1063/1.472798,10.1063/1.473231, 10.1063/1.480028,10.1063/1.1703704}, but they suffer from the so-called sign problem\cite{Alexandru_2022}. As a result, these methods are applicable to cases where only when few degrees of freedom interact\cite{muhlbacher_real-time_2008} and cannot address realistic quantum dynamics of complex systems. This is due to the difficulty in converging calculations with respect to the number of paths of opposite signs, which exponentially increases with the number of degrees of freedom. An additional problem with such approaches is the difficulty in capturing quantum tunneling phenomena, which is especially important for the dynamics of STO. Imaginary path integral methods avoid some of these problems but are only rigorously formulated for equilibrium statistical properties and cannot address transient light-induced response phenomena. \cite{10.1063/1.471221, RevModPhys.67.279,10.1063/1.441588,10.1063/1.1777575,10.1063/1.2074967} Furthermore, their computational cost diverges as the temperature approaches absolute zero. 
Therefore, there is a need for an efficient non-perturbative simulation method applicable at low temperatures and capable of capturing real-time out-of-equilibrium quantum dynamics of the nuclei.

In this work, we introduce a new methodology for efficiently solving the quantum nuclear Schr\"odinger equation and compute the nonlinear dynamics of STO under intense pulsed THz radiation from first principles without any assumption on the phonon interactions. This is achieved by formulating a rigorous time-integration scheme \cite{libbi2024atomisticsimulationsoutofequilibriumquantum} for calculating the quantum nuclear dynamics through the time-dependent self-consistent harmonic approximation (TDSCHA) \cite{PhysRevB.103.104305, PhysRevB.107.174307} combined with an efficient description of the Born-Oppenheimer surface using a state-of-the-art machine-learning force field\cite{Vandermause2022_,Xie2021} trained on density functional theory (DFT).

We start the simulation at time $t=\SI{-1000}{\femto\second}$ with a supercell of 40 atoms of STO thermally equilibrated at \SI{100}{\kelvin}. This initial state is achieved via the stochastic self-consistent harmonic approximation (SSCHA)\cite{Monacelli_2021}, an equilibrium approach already validated for correctly capturing the quantum suppression of ferroelectricity in STO\cite{PhysRevMaterials.7.L030801}. Then, at $t=\SI{0}{\femto\second}$, the system interacts with a single-cycle electric field in resonance with the SPM (the THz laser pump, shown in \figurename~\ref{fig1}\textcolor{blue}{a}) and is driven out of equilibrium.  The ability to generate such a single-cycle oscillation of the electric field has only recently become possible due to advances in control of THz fields \cite{Kozina2019}. The dynamical coupling between the laser light and the nuclear coordinates is determined by the Born effective charges, evaluated within DFT at the equilibrium structure and assumed constant throughout the simulation. The quantum nuclear density matrix is evolved according to forces obtained from the machine learning interatomic potential\cite{Vandermause2022_, Xie2021} (more details in the Methods section) and additional force contributions due to light-matter interaction for the first few hundreds of femtoseconds (pulse duration).
To mirror the experimental conditions documented in a prior study by Kozina et al. \cite{Kozina2019}, the electric field was oriented along the [1,-1,0] crystallographic direction of STO's reference cubic cell. We set the field's maximum amplitude to 833 kV/cm.

At \SI{100}{\kelvin}, the primitive cell is nearly cubic (the antiferrodistortive transition occurs at 105K), wherein the phonons at the $\Gamma$ point are organized into four distinct groups of degenerate modes, represented in \figurename~\ref{fig1}\textcolor{blue}{c}. 

The first group (\TOo) includes the soft phonon modes (SPMs) with a frequency of \SI{1.5}THz, which are responsible for the ferroelectric transition when nuclear quantum fluctuations are suppressed (e.g. through $\mathrm{^{18}O}$ substitution \cite{PhysRevLett.96.227602}). The remaining groups of degenerate phonons have frequencies \SI{4.8}{\tera\hertz}, \SI{7.7}{\tera\hertz}, and \SI{15.4}{\tera\hertz}. The phonon modes polarized along the direction of the electric field are labeled respectively as \TOd, \TOt, and \TOf.

Shortly after the pump interacts with the system and the SPM is excited, the phonon-phonon interactions transfer the energy in a complex way between all 117 phonon modes compatible with our simulation cell. 
To examine how energy flows from the excited SPM to the other degrees of freedom, we decomposed the total energy into the time-dependent contributions of individual phonons, reported in \figurename~\ref{fig1}\textcolor{blue}{b} (implementation details are in the Methods). Approximately \SI{300}{\femto\second} after the pump arrives, the \TOd, \TOt, and \TOf modes begin oscillating, as shown in \figurename~\ref{fig1}\textcolor{blue}{d}. This process has been observed for the \TOd~and \TOt~modes by time-resolved X-ray diffraction\cite{Kozina2019}, and is referred to as phonon upconversion. 

After \SI{1.0}{\pico\second}, the anharmonic phonon scattering transfers energy to modes at finite momenta. We observe a background of almost all modes that gradually acquire energy, damping the motion generated by the pulse. These processes in STO have never been identified so far, neither from experiment nor theoretical simulations, as they arise from a multiple interaction of phonons modes after all the $\Gamma$ bands have been excited, testifying to the importance of the atomistic description to capture the out-of-equilibrium quantum dynamics of nuclei. 

To gain a better insight into the process of the upconversion we calculate the Fourier transform of the $\mathrm{\Gamma}$-modes oscillations. As depicted in \figurename~\ref{fig1}\textcolor{blue}{e}, the vibrational spectrum of modes \TOd, \TOt, and \TOf peaks at their natural frequency. However, the SPM oscillates at a higher frequency (approximately \SI{2.0}{\tera\hertz} instead of its natural frequency of \SI{1.5}{\tera\hertz}) due to the stiffening caused by the anharmonic terms in the PES of the SPM when driven to large oscillations by large electric fields. Such stiffening as a function of the amplitude of the electric field is illustrated in \figurename~\ref{fig1}\textcolor{blue}{f}. The broadening of the spectrum with increasing pulse intensity is associated with a reduction in the SPM lifetime due to the increased energy transfer from the SPM to the phonon bath, which will be analysed in more detailed later in the main text.   

\begin{figure*}[t]
    \centering
    \includegraphics[width=\textwidth]{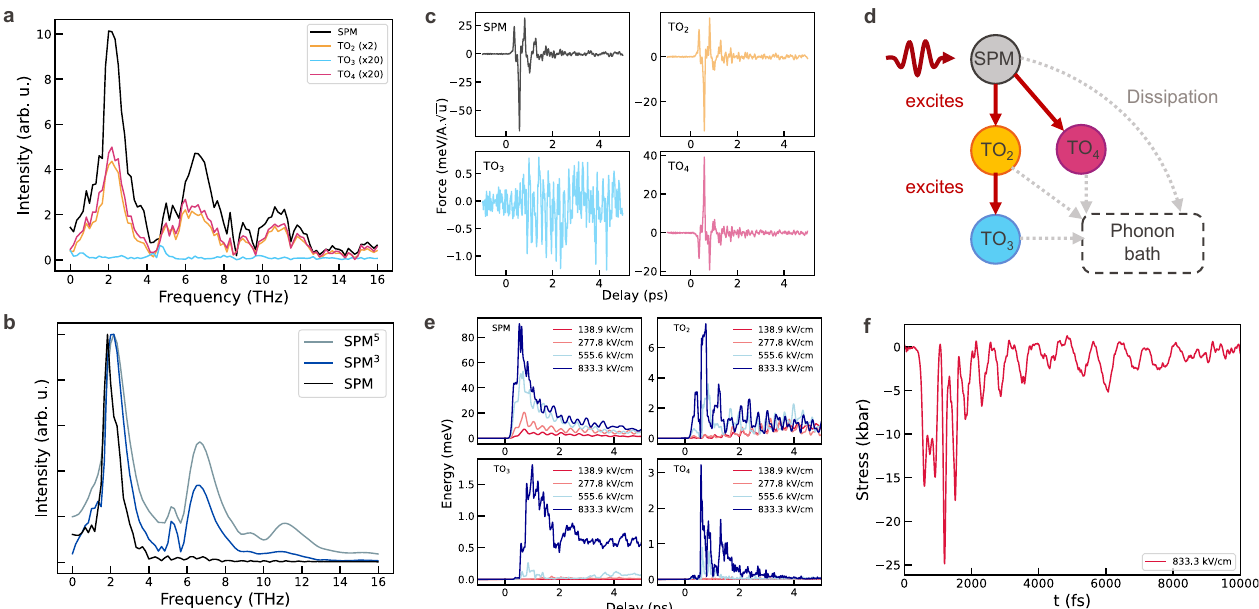}
    \caption{\textbf{a}, Fourier transform of the anharmonic force projected on the phonon modes at $\mathrm{\Gamma}$.  \textbf{b}, Fourier transform of the coordinate of the SPM mode, raised to different powers. \textbf{c}, Time-dependent anharmonic force projected on the phonon modes at $\mathrm{\Gamma}$. \textbf{d}, Schematic representation of the energy flow between phonon modes. \textbf{e}, Energy of the $\mathrm{\Gamma}$-phonons as a function of the amplitude of the electric field. \textbf{f}, The stress induced on STO by the irradiation with a THz pulse of amplitude 833.3 kV/cm.}
    \label{fig2}
\end{figure*}

To unveil the mechanism of the excitation of the \TOd, \TOt, and \TOf modes, we plot in \figurename~\ref{fig2}\textcolor{blue}{a} the Fourier transform of the average anharmonic force projected on the atomic motion of each mode $\mu$. This is obtained by subtracting the harmonic restoring force $-\omega_{\mu}^2Q_{\mu}$ from the total average force acting on the phonon mode. Here $\omega_{\mu}$ refers to the phonon natural frequency, while $Q_{\mu}$ to its displacement amplitude (details in the Methods section). The anharmonic force spectrum for the modes SPM, \TOd\ and \TOf is characterized by three prominent peaks with amplitudes decreasing linearly as a function of frequency. The ratios of the peak heights matches very well with that of the 5th power of $Q_{SPM}$, which is illustrated in \figurename~\ref{fig2}\textcolor{blue}{b}. This suggests that high electric fields (833 kV/cm) excited the SPM sufficiently strongly to drive the system far from equilibrium into a highly anharmonic part of the free energy surface. 
A much different case is that of the \TOt\ mode, whose anharmonic force spectrum is peaked at the \TOd frequency, suggesting that it is excited through the \TOd mode. The plot of such forces in time (\figurename~\ref{fig2}\textcolor{blue}{c}) shows that the sudden oscillation of the SPM far from equilibrium generates an impulsive force on the \TOd and \TOf modes through anharmonic coupling. The coherent oscillations of \TOd then drive the \TOt mode (see \figurename~\ref{fig2}\textcolor{blue}{d} for a schematic representation). 

To further delve into the process of the phonon upconversion, we analyze the response of the $\Gamma$ phonon modes while varying the amplitude and the frequency of the pump pulse.
\figurename~\ref{fig2}\textcolor{blue}{e} illustrates the evolution of the energy of these modes as a function of the pulse amplitude when it is in resonance with the frequency of the SPM mode at \SI{100}{\kelvin}. The responses of these modes increase non-linearly with the amplitude of the electric field. The energy of oscillation of the higher-frequency TO modes correlates with that of the SPM. Notably, we observe almost complete suppression of phonon upconversion for amplitudes below 277.8 kV/cm, further indicating the nonlinear character of this process. Moreover, the full atomistic descriptions of the system enable us to access the lifetime of the phonon mode excitation, as the large number of degrees of freedom acts as the thermal reservoir into which the energy of the excited subsystem eventually decays. In particular, our simulations did not employ any empirical smearing or dephasing factor, thus providing the real lifetime arising from the natural anharmonic phonon scattering among the atomic degrees of freedom. The lifetime of the SPM is strongly affected by the pump intensity. Previous experimental results showed that for applied pulses with amplitude below \SI{80}{\kilo\volt\per\centi\metre}, the SPM lifetime increases with the laser intensity\cite{katayama_ferroelectric_2012}, manifested as a spectral narrowing. However, the onset of phonon upconversion at amplitudes around \SI{280}{\kilo\volt\per\centi\meter} points to the appearance of new pathways for energy to decay. The SPM lifetime further decreases upon increasing pump amplitude, thus unveiling a complex dependence of the transient dynamics lifetimes with the pulse intensity. The ability of our explicit fully atomistic dynamics simulation to capture amplitude-dependent damping of the SPM mode goes beyond previous models that assumed an effective constant damping factor\cite{fechner_quenched_2024,PhysRevLett.129.167401,Kozina2019,PhysRevLett.130.126902, PhysRevX.11.021067}.

\begin{figure}[h]
    \centering
    \includegraphics[scale=0.5]{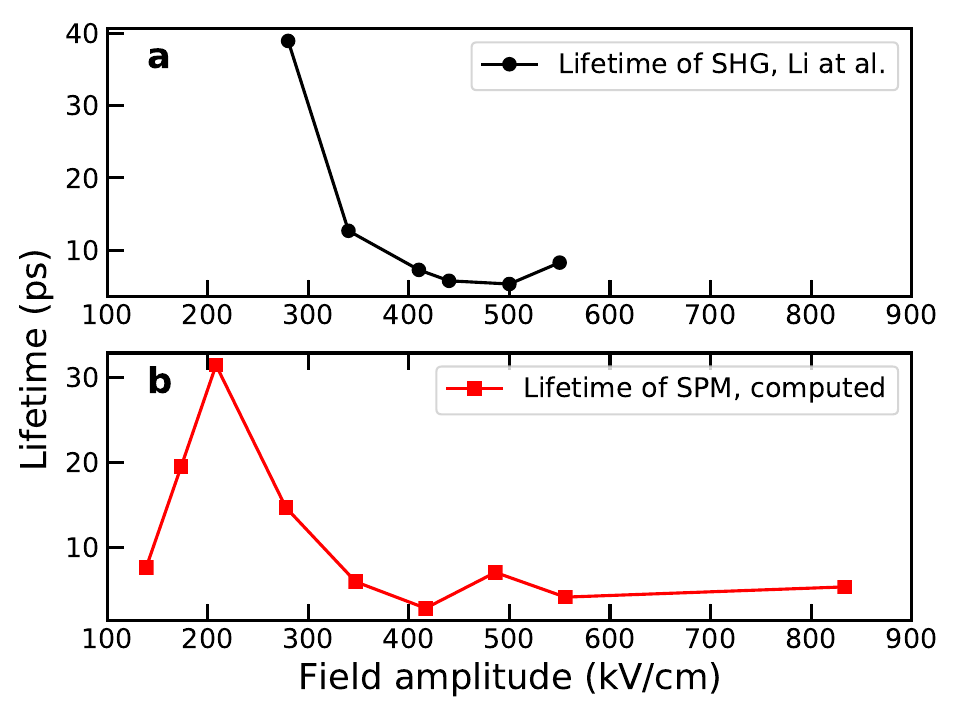}
    \caption{\textbf{a}, lifetime of the SHG extracted from the measurements in Ref. \cite{doi:10.1126/science.aaw4913}. \textbf{b} lifetime of the SPM calculated at 100~K. }
    \label{lifetimes}
\end{figure}

There is currently a vigorous debate about the origin of the SHG signal observed in the experiments of Ref. \cite{doi:10.1126/science.aaw4911, doi:10.1126/science.aaw4913}. The theoretical work of Ref. \cite{PhysRevLett.129.167401} suggests that this is consistent with the presence of a metastable ferroelectric state, which requires the interaction between the SPM with other two lattice modes outside $\Gamma$. This state is reached by the system during its evolution in response to the terahertz (THz) pulse. In Refs. \cite{doi:10.1126/science.aaw4911,PhysRevLett.129.167401}, lattice strain has been identified as the factor responsible for such a ferroelectric transition. STO in equilibrium at low temperature is paraelectric but very close to a ferroelectric instability, with the paraelectric structure stabilized by the quantum zero-point motion of nuclei\cite{PhysRevMaterials.7.L030801}. A positive strain increases the barrier between the two minima in STO's potential energy surface (PES), thereby potentially stabilizing the ferroelectric phase. To determine whether such a metastable state exists, we conduct SSCHA relaxations to identify free energy minima of the 20-atom STO cell over a range of isotropic strains from $\mathrm{-0.3}$\% to 0.9\% at 0K (see \figurename~S2 in the SM). We observed that a strain of as 0.6\%, corresponding to a stress of 2.3 GPa, is sufficient to destabilize the paraelectric phase and induce a ferroelectric transition. These values are overestimated because the PBE functional underestimates the height of the energy barrier. They can be corrected using a 1D model tuned on RPA calculations, yielding a transition strain of 0.1\% and a pressure as small as few hundred kbar. 
Details on the calculation are in the Supplementary Information (SI) \cite{supplementary}). When the strain is increased further, 
we observed a monotonic increase of the free energy, indicating that no metastable state exists with a simple ferroelectric order with periodicity in the 20-atom cell. This implies that a static metastable ferroelectric distortion, if any, may occur on a larger periodicity. A dynamical ferroelectric transition remains possible, induced by the non-equilibrium transient strain resulting from the system's dynamic response to a light pulse. 
To verify this occurrence, we perform TDSCHA simulations where the system is prepared at \SI{0}{\kelvin} and pumped with a pulse polarized along the c-axis with maximum field of \SI{833}{\kilo\volt\per\centi\metre}. 
Our TDSCHA approach at this point is only applicable at fixed volume and does not allow us to explore dynamic strain effects and possible ferroelectric transitions, unless extremely large cells are adopted.
However, we can analyze the instantaneous stress resulting on the cell, accounting for the quantum and thermal fluctuations of the out-of-equilibrium dynamics using the formalism introduced in Ref.~\cite{PhysRevB.98.024106}. 
When the pump interacts with the sample, a massive stress \SI{2.5}{\giga\pascal} is induced along the c-axis. Intriguingly, after the end of the pump, the stress persists and oscillates around negative values (corresponding to tensile stress in our convention), at a frequency that is twice that of the SPM at 0K, which ranges from 0.6 THz to 1.0 THz (because of the stiffening due to anharmonicity, as discussed above). This residual stress is significant, with peak values reaching 0.5 GPa.
If this stress were to induce a strain of the cell, it would be sufficient to generate a ferroelectric dynamical phase transition, where the structure's oscillations around the local minima of the free energy could persist as long as the atomic motion has enough energy to sustain the strain in the system. The propagation of strain would occur as acoustic waves within the material. In a perfectly bulk material, the time required for the strain to manifest is excessively long, incompatible with the picosecond time scale of the transition observed in Ref. \cite{doi:10.1126/science.aaw4913}. However, the presence of polar nano-regions stemming from defects in STO\cite{PhysRevLett.130.126902}, spaced at an average distance of nanometers, could drastically reduce this time, thereby rendering it compatible with a picosecond time scale. Specifically, this time corresponds to the duration needed for an acoustic wave with a group velocity of approximately ~$\mathrm{10^3}$ m/s to traverse nanometer-scale boundaries.
A similar mechanism was proposed by Shin et al. \cite{PhysRevLett.129.167401}, where, however, the presence of few degrees of freedom makes the ferroelectric state metastable, thus resulting in a persisting SHG signal. This did not allow the comparison between the lifetime of the ferroelectric state with the available experimental data. 
Furthermore, we observe that, due to the increase of the anharmonic phonon scattering above 200 kV/cm, a more intense pump field results in a shorter lifetime of the SPM and, thus, of the dynamical stress. In \figurename~\ref{lifetimes}, we compare the simulated lifetime of the SPM as a function of the electric field amplitude at \SI{100}{\kelvin}, which is a good probe of the duration of the transient nonlinear dynamics, with the measured lifetime of the inversion symmetry breaking observed by SHG after THz pulse at low temperature\cite{doi:10.1126/science.aaw4913}.

Both lifetimes show a strong correlation above \SI{270}{\kilo\volt\per\centi\meter}, the field value corresponding to the onset of both the phonon upconversion and the inversion symmetry breaking. Interestingly, the lifetimes of both the SHG and the SPM decrease with increasing field amplitude. This is the opposite of the trend reported by Cheng et al. \cite{PhysRevLett.130.126902} for the \ch{KTaO3} perovskite, where the inversion symmetry breaking was associated with THz-driven dipole correlations of defect-induced local polar structures. On the other hand, in STO, the strong correlation of the SHG signal with the lifetime of the SPM suggests that the SHG signal is brought about by complex out-of-equilibrium lattice dynamics induced by the electric field pulse. In other words, inversion symmetry breaking does not correspond to the presence of a metastable or thermodynamically stable phase in STO, as originally hypothesized\cite{doi:10.1126/science.aaw4913}. Instead, ferroelectricity only exists as a dynamical transient out-of-equilibrium state that can persist only as long as the atomic motion maintains enough stress to deform the local potential into a sufficiently deep double well, allowing the spontaneous breaking of inversion symmetry.
In conclusion, thanks to a novel framework for simulating the out-of-equilibrium quantum nuclear dynamics from the first principles, we unveil the origin of the complex nonlinear phenomena observed in STO under intense THz laser pulses.
We disclose the microscopic mechanism of the phonon upconversion and the complete energy decay pathway across different phonons, both coherent and incoherent, following laser excitation. Thanks to a parameter-free description, we identified a fingerprint of the ferroelectric transition in the light-induced stress calculated in STO, substantial enough to induce a polar order.

Our findings pave the way for a fully first-principles investigation and identification of new materials that can host light-induced phase transitions.

We acknowledge useful discussions with Zhuquan Zhang. 
This research was funded in part by the Swiss National Science Foundation (SNSF, Mobility fellowship P500PT\_217861), the Department of Navy award N00014-20-1-2418 issued by the Office of Naval Research and Robert Bosch LLC.
L. M. thanks the European Union under the program Horizon 2020 for the award and funding of the MSCA individual fellowship (grant number 101018714)
Computational resources were provided by the FAS Division of Science Research Computing Group at Harvard University.

\section{Methods}
The nonlinear quantum dynamics of nuclei is simulated through the time-dependent self-consistent harmonic approximation (TDSCHA)\cite{PhysRevB.103.104305}.
The TDSCHA assumes that the nuclear quantum density matrix is Gaussian in the Wigner phase-space \cite{PhysRev.40.749,10.1063/1.1705323}, parameterized by the mass-rescaled position-position,  momentum-momentum, and position-momentum covariances $\bA=\braket{\delta \bR \delta \bR}$, $\bB=\braket{\delta \mathbf{P} \delta \mathbf{P}}$, $\bG=\braket{\delta \bR \delta \mathbf{P}}$ and centered at the average atomic positions $\mR$ and momenta $\mP$.
TDSCHA represents the dynamic generalization of the SSCHA, which has proven to describe very well the vibrational and structural properties of different categories of anharmonic materials and their temperature dependence \cite{Monacelli_2021,Errea2020,PhysRevLett.125.106101,PhysRevLett.122.075901}. Particularly, SSCHA reproduces the structural properties of perovskites very accurately\cite{Monacelli2023} and predicts the correct quantum paraelectric phase for STO \cite{PhysRevMaterials.7.L030801}. Furthermore, the linear response of TDSCHA has allowed for an unprecedented description of the Raman and infrared spectra of metallic hydrogen \cite{Monacelli_Nature_2021}.
The dynamics for the parameters $\mR$, $\mP$, $\bA$, $\bB$ and $\bG$ is obtained by imposing the least action principle \cite{PhysRevB.107.174307}, leading to the self-consistent time-dependent equation for the density matrix:
\begin{equation}
    i\hbar\frac{\partial\hat\rho}{\partial t} = \bigl[\mathcal H[\hat\rho], \hat\rho \bigr],
\end{equation}
where $\hat\rho$ is the nuclear quantum density matrix, $\mathcal H[\hat \rho]$ is the self-consistent harmonic Hamiltonian whose parameters depend on the anharmonic potential and the density matrix $\hat\rho$ at the same time, and the square brackets indicate the quantum commutator\cite{PhysRevB.103.104305}.
Substituting the Gaussian expression for the density matrix leads to the set of differential equations
\begin{equation}\label{tdscha}
\begin{cases}
    \dot{\mR} = \mP \\
    \dot{\mP} = \braket{\mathbf{f}} \\
    \dot{\bA} = \bG + \bG^{\dag} \\
    \dot{\bB} =  -\dV\bG - \bG^{\dag}\dV  \\
    \dot{\bG} =   \bB-\bA\dV \ .
\end{cases}
\end{equation}
We assume that the THz pulse does not excite electrons to higher energy states, as the bandgap of STO (3.2 eV) is significantly larger than the THz photon energy. Consequently, the ionic motion is described by the ground-state Born-Oppenheimer potential energy surface.\\
The dynamics simulations are significantly accelerated using a machine-learned interatomic potential (MLIP) trained with state-of-the-art active learning strategies. This active learning is conducted over a run of 1000 ps, covering a range of strains from -2\% to 2\% and temperatures up to 500K.\\
The potential is trained on DFT calculations with the PBE functional, as described in detail in the first section of SI.
According to the work of Verdi et al. \cite{PhysRevMaterials.7.L030801}, this results in an underestimate of the barrier height of the potential energy surface of the SPM (third section of SI). Such underestimation leads to a increase of the frequency of the ferroelectric soft mode, which in our calculations is $\nu=2.6$ THz at 100K and $\nu=1.5$ THz at 0K, compared to the experimentally determined $\nu\sim1.5$ THz at 100 K \cite{Kozina2019} and $\nu\sim0.5$  THz at 0K \cite{Yamanaka_2000}. To address this limitation, we apply a uniform strain to the STO cell. The strain value is chosen to match the experimental frequency of the SPM as obtained from SSCHA linear-response calculations. The adopted strain values are then 0.3\% at 100K and 0.5\% at 0K. These very small strains are expected to leave the coupling constant between phonon modes unchanged, as we tested by observing negligible frequency changes in all other modes.\\
The pulse shape used to study the phonon upconversion is modelled as 
\begin{equation}
f(t) = -A\ \frac{t}{\sigma}e^{-\frac{t^2}{2\sigma^2}+\frac{1}{2}}\ ,
\end{equation}
with $\sigma=1/(2\pi f_{SPM})\simeq$ 0.1 ps (more details in the SI). 

To analyze the energy transfer between different vibrational modes, we project the atomic motion onto the SSCHA-computed phonons at 100K using the following relations
\begin{equation}
    Q_{\mu} = \sum_{a} \sqrt{m}_a e_{\mu a} u_a\ , 
\end{equation}
where $\mu$ is the phonon branch index, $a$ labels both atoms in the supercell and Cartesian coordinates, $e_{\mu a}$ is the phonon eigenvector at equilibrium and $u_a=\mathcal{R}_a(t)-\mathcal{R}_a(0)$ represents the centroid displacement from the equilibrium position . We define $h_\mu(t)$ as the out-of-equilibrium local energy 
of the $\mu$-th phonon
\begin{equation}
    h_{\mu}(t) = \frac{\dot{Q}_{\mu}^2}{2} + \frac{\omega_{\mu}^2Q_{\mu}^2}{2}\ ,
\end{equation}
where the two terms in the r.h.s. represent the phonon kinetic energy and harmonic elastic energy, respectively. The sum of $h_\mu$ on all modes is the centroid's energy. To obtain the system's total energy, we  also consider the effects of the quantum-thermal dispersion and the anharmonicity, which cannot be repartitioned on a mode-by-mode basis. However, these contributions are almost constant during the motion; thus, $h_\mu(t)$ is a good representation of the energetic contribution of each mode.
The energy spectral density, 
\begin{equation}
    h(\omega,t) = \sum_{\mu}h_{\mu}(t)\delta(\omega-\omega_{\mu})\ ,
\end{equation}
is shown in Fig. \ref{fig1}\textcolor{blue}{b}. The area outlined in red indicates the spectrogram of the terahertz pump pulse (contour at 30\% of peak magnitude).

The average force acting on the mode $\mu$ can be calculated as
\begin{equation}
    \braket{f}_{\mu} = \sum_i\frac{\braket{f}_i}{\sqrt{m_i}}e_{\mu i}\ .
\end{equation}
The expected anharmonic force is obtained as
\begin{equation}
    f_{\mu}^{anh} = \braket{f}_{\mu} + \omega_{\mu}^2Q_{\mu}\ ,
\end{equation}
the first term is the total force acting on the mode $\mu$, and the second subtracts the harmonic force. $f^{anh}_{\mu}$ is the quantity represented in \figurename~\ref{fig2}\textcolor{blue}{c} of the main text.

The effective charges describe the interaction between laser light and nuclear coordinates, accounting for the dielectric screening of local fields, as discussed in the second section of SI.


\bibliographystyle{ieeetr}

\end{document}